\documentclass[a4paper]{article}

\usepackage{icrc2013}

\title{VERITAS Observations of TeV Binaries}

\shorttitle{VERITAS Observations of TeV Binaries}

\authors{
Andrew W. Smith$^{1}$,
for the VERITAS Collaboration.
}

\afiliations{
$^1$ University of Utah Physics and Astronomy \\
}

\email{aw.smith@utah.edu}

\abstract{Since the commissioning of the array in Spring 2007, the VERITAS array (sensitive in
the 0.1-50 TeV energy range) has acquired over 300 hours of observations investigating
the TeV emission from X-ray binary star systems, in particular focusing on the known
TeV binary targets LS I +61$^{\circ}$ 303 and HESS J0632+057.  Both TeV binaries have
been monitored by VERITAS for several years and the resulting dataset is continuing to
yield important results in the characterization of these poorly understood systems. We
present these results, as well as the contemporaneous observations of these sources
taken with $Fermi$-LAT and $Swift$-XRT.  In the case of LS I +61$^{\circ}$ 303, simultaneous
observations taken with VERITAS and $Fermi$-LAT reveal a distinct ÒgapÓ in emission in
the 10-200 GeV range. For HESS J0632+057, the extended VERITAS observations have allowed for the first identification of a binary system through TeV gamma-ray observations.}


\begin{document}
\maketitle

%Begin a section.
\section{VERITAS and TeV Binaries}
The VERITAS array \cite{bib:VERITAS} of imaging atmospheric Cherenkov telescopes (IACTs) located in southern Arizona (1.3 km m.a.s.l., 31$^{\circ}$40'30''N, 110$^{\circ}$57'07'' W) began 4-telescope observations in September 2007. The array is composed of four 12m diameter telescopes, each with a Davies-Cotton tessellated mirror structure of 345 12m focal length hexagonal mirror facets (total mirror area of 110 m$^{2}$). Each telescope focuses Cherenkov light from particle showers onto its 499-pixel PMT  (photomultiplier tube) camera. Each pixel has a field of view of 0.15$^{\circ}$, resulting in a camera field of view of 3.5$^{\circ}$. VERITAS has the capability to detect and measure gamma rays in the 85 GeV to 50 TeV energy regime with an energy resolution of 15-20$\%$, an angular resolution of $<$0.1$^{\circ}$ on an event by event basis, and a flux sensitivity of 1$\%$ Crab Nebula in a $\sim$ 25 hour observation.

The current generation of IACT instruments  such as VERITAS, HESS \cite{bib:HESS}, and MAGIC \cite{bib:MAGIC} have not only greatly expanded the study of  previously known TeV sources, such as active galactic nuclei, pulsar wind nebula, and radio galaxies, they have also provided several new classes of objects for study in the TeV regime. Among these is the class of X-ray binary star systems which also present strong, episodic flaring in the TeV regime. The study of these systems has been a high priority for VERITAS since the inception of the array in 2007, with over 300 hours of lifetime observations accrued on a wide variety of X-ray binary systems believed to be good candidates for TeV emission. The observations have ranged from multi-wavelength monitoring campaigns on the known TeV binaries such as LS I +61$^{\circ}$ 303 \cite{bib:USLSI1,bib:USLSI2,bib:USLSI3,bib:USLSI4} and HESS J0632+057 \cite{bib:GernotHESS,bib:GernotNext}, to target of opportunity (ToO) observations of TeV binary candidates triggered by activity at other wavelengths on sources such as Cyg X-3 \cite{bib:Angelo}, 1A05353+262 \cite{bib:1A0535}, and V407 Cygni \cite{bib:V407}.

Before observations of these systems commenced with the latest generation of IACTs, X-ray binary targets such as LS I +61$^{\circ}$ 303, LS 5039. Cygnus X-3, Cygnus X-1, GRS 1915+105, SS 433, and PSR B1259-63 were all believed to be excellent candidates for TeV emission due to their strong non-thermal X-ray flux, radio flares, and a wide range of evidence for relativistic particle acceleration taking place. However, after 8 years of observations at the $\sim$2$\%$ Crab flux sensitivity level,  only LS I +61$^{\circ}$ 303, LS 5039, and PSR B159-63 have been reliably detected at TeV energies. The important question then seems to be: why only these systems? Furthermore, the discovery of HESS J0632+057 in TeV gamma-ray \textit{first}, with its association as an X-ray binary only coming afterwards, makes the characterization of the TeV behavior from X-ray binaries even more convoluted. 

 Due to its Northern Hemisphere location, VERITAS has been able to provide deep observations of both LS I +61$^{\circ}$ 303 and HESS J0632+057. These observations, along with data taken by multi-wavelength partners (such as $Swift$-XRT and $Fermi$-LAT), have yielded a wealth of information that, in some ways, have provided some insight about the physical nature of these systems; in other cases the increase dataset has confused the situation even further. Here we report on the observation of LS I +61$^{\circ}$ 303 and HESS J0632+057 taken with VERITAS over the last several years.
 
 \section{LS I +61$^{\circ}$ 303}
 
 LS I +61$^{\circ}$ 303 is a high mass X-ray binary star system (HMXB) known to be the pairing of a compact object (either black hole or neutron star) and a large (Be or O star) main sequence companion, located at a distance of 2 kpc \cite{bib:HandC, bib:Casares2005}. Radial velocity measurements show the orbit to be elliptical ($e=0.537\pm0.034$), with periastron passage occurring around phase $\phi=0.275$, apastron passage at $\phi=0.775$, superior conjunction at $\phi=0.081$ and inferior conjunction at $\phi=0.313$ \cite{bib:Aragona2009}. LS I +61$^{\circ}$ 303 is known for its energetic outburst across the electromagnetic spectrum (including IR, radio, X-ray, optical, and both GeV and TeV gamma ray), modulated with the orbital period of the compact object ($\sim$26.5 days). LS I +61$^{\circ}$ 303, first detected as a TeV source by MAGIC in 2005 \cite{bib:MAGICdetection}, has been observed extensively by VERITAS since 2006.  VERITAS has now accrued over 150 hours of quality selected lifetime observations of LS I +61$^{\circ}$ 303, and has provided in-depth characterization of its TeV behavior available. The source is a variable TeV emitter, showing emission at the 2-20$\%$ Crab flux level above 200 GeV, typically near the apastron passage ($\sim\phi$=0.5-0.8) of the compact object. The timing of the TeV emission is poorly understood: while first detected only near apastron passage in 2005-2007, the source did not show any clear TeV emission in observations taken from 2008-mid 2010. However, in October 2010, the source was re-detected by VERITAS, this time near periastron passage \cite{bib:USLSI3}. Since then, LS I +61$^{\circ}$ 303 has been detected near apastron passage again by both VERITAS and MAGIC, appearing to have returned to behavior seen in 2005-2007 \cite{bib:USLSI4, bib:MAGICLOW}. It is possible that the emission from the system may go through a complex, multi-year modulation, similar to that seen in radio \cite{bib:Gregory2002} and VERITAS is dedicated to accruing a continued dataset in order to examine this behavior.
 
 \begin{figure}[t]
  \centering
  \includegraphics[height= 0.32\textwidth,width=0.5\textwidth]{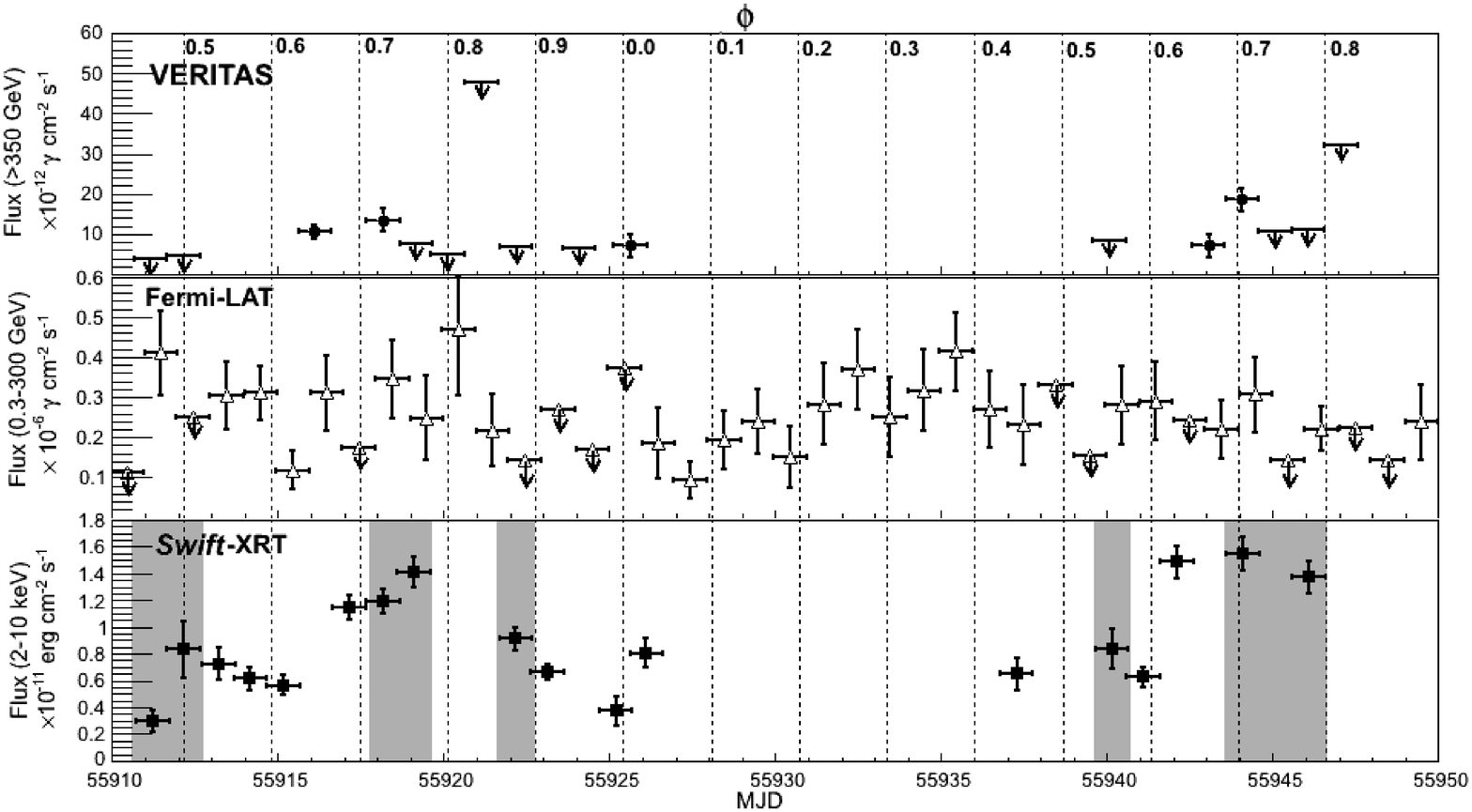}
  \caption{The VERITAS ($>$350 GeV daily integrations, top), Fermi-LAT (0.3-300 GeV, middle), and Swift-$\textit{XRT}$ (0.3-10 keV, bottom) light curves for LS I +61$^{\circ}$ 303 during December 2011 - February 2012. The data is also shown as a function of orbital phase ($\phi$). VERITAS 99$\%$ and $Fermi$-LAT 90$\%$ flux upper limits are represented by arrows. The grey shaded regions represent the observations obtained simultaneously which are used for the X-ray/TeV correlation studies in this work. }
  \label{Fig1}
 \end{figure}

 From December 2011-February 2012, VERITAS observed LS I +61$^{\circ}$ 303 as part of an extensive multi-wavelength campaign utilizing observations with $\textit{Swift}$-XRT (2-10 keV) and $Fermi$-LAT (0.3-300 GeV). The source was observed for a total of 25 hours of quality selected livetime with VERITAS, resulting in a TeV detection near apastron passage at a 11.9$\sigma$ statistical significance.  This detection occurred near apastron passage of the compact object in its orbit, with the source presenting a flux of 5-15 $\times 10^{-12} \gamma$s cm$^{-2}$s$^{-1}$ above 350 GeV, or approximately 5-15$\%$ of the Crab Nebula flux in the same energy regime. 
 
 While flux levels of this kind (and at this orbital phase) have been seen before in LS I +61$^{\circ}$ 303, for the first time these observations reveal evidence for nightly variability in the source (see Figure 1). The observations taken on MJD 55918/55919 and MJD 55944/55945 show evidence for a flux decrease at the 2.7$\sigma$ and 3.6$\sigma$ significance level respectively. These significances are post-trials, accounting for nine trials (one trial for each nightly pair tested).  The flux differences seen only present evidence for the TeV flux falling off on a nightly timescale and do not necessarily imply that the TeV flux from LS I +61$^{\circ}$ 303 increases on such a rapid timescale.  
 
 On eight separate observations, VERITAS was able to obtain data simultaneously with \textit{Swift}-XRT observations in the 2-10 keV energy range (see Figure 2). Analysis of these observations shows no strong evidence for correlation with a Pearson correlation coefficient of 0.36$\pm$0.32, consistent with two uncorrelated datasets. The contemporaneous observations (daily fluxes) taken with VERITAS and $Fermi$-LAT do not show any strong evidence for correlation either, with a derived correlation coefficient of 0.1$\pm$0.3.
  \begin{figure}[t]
  \centering
  \includegraphics[width=0.5\textwidth]{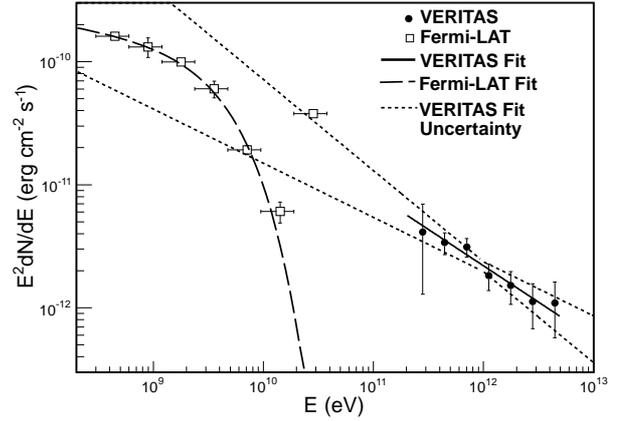}
  \caption{The spectral energy distribution of LS I +61$^{\circ}$ 303 derived from contemporaneous $Fermi$-LAT and VERITAS observations in 2011/2012 showing a clear gap in emission between $\sim$4 GeV and 300 GeV.}
  \label{Fig3}
 \end{figure}

      \begin{figure}[h]
  \centering
  \includegraphics[width=0.4\textwidth,height=0.2\textheight]{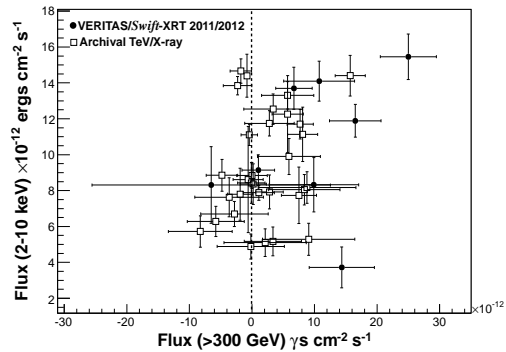}
  \caption{The strictly simultaneous TeV/X-ray observations taken on LS I$^{\circ}$. Filled circles represent data from this work and the open squares represent observations detailed in \cite{bib:USLSI3} and references therein. The total correlation between the flux in each band is 0.36$\pm$0.32, consistent with two uncorrelated datasets.}
  \label{Fig2}
 \end{figure}

 Additionally, the observations taken by VERITAS in 2011/2012 allow for the construction of a GeV-TeV energy spectrum (Figure 3). The TeV spectrum is well fit by a power-law described by $(1.37\pm0.14_{stat})$$\times$10$^{-8}$ $\times$($\frac{E}{\mathrm{1 TeV}})^{-2.59\pm0.15_{stat}}$ TeV$^{-1}$ m$^{-2}$ s$^{-1}$. Comparing this spectrum to the energy spectrum derived from the GeV observations taken by $Fermi$-LAT during the same time period as the VERITAS detection shows a power law with an exponential cutoff at $\sim$ 4 GeV, showing a distinct gap in emission between 4 GeV and 350 GeV, although the VERITAS energy spectrum does not extend low enough to determine the exact extent of the gap. It is possible that that this gap is indicative of the multiple particle populations at work in LS I +61$^{\circ}$ 303, which would also explain the apparent lack of correlation between past GeV and TeV observations.  It also possible that the GeV spectrum undergoes some fast variability connecting to it to the TeV spectral points, however, such fast GeV spectral variability has not been observed in LS I +61$^{\circ}$ 303 since the launch of $Fermi$-LAT.

 Further observations to constrain the low energy cutoff of the TeV spectrum are needed to examine this phenomena. The recent VERITAS upgrade incorporating high quantum efficiency photomultiplier tubes will aid in this regard, significantly lowering the energy threshold of observations of LSI +61 303 \cite{bib:Kieda}.  It is clear that the existence of day-scale variability as well as the apparent separate production of the GeV and TeV emission will effect modeling of this system. For instance, it is striking that the GeV spectrum of LS I +61$^{\circ}$ 303 is strongly reminiscent of the spectra of many $Fermi$-LAT pulsars and may indicate that the emission is powered by a young, energetic pulsar. Please see \cite{bib:USLSI4} for additional details.

     \begin{figure}[t]
  \centering
  \includegraphics[width=0.52\textwidth]{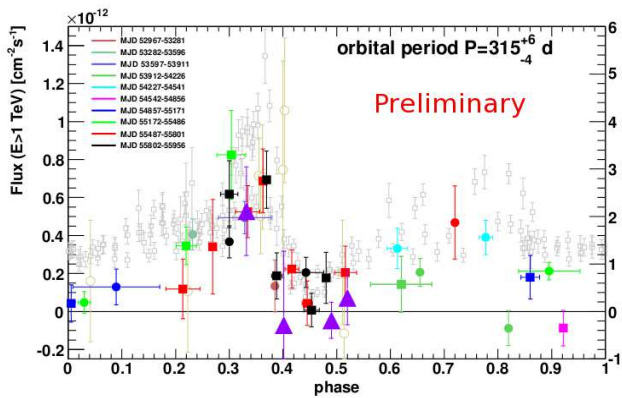}
  \caption{The phase folded TeV light curve of HESS J0632+057 (color points) from VERITAS (squares and triangles), HESS (filled circles), and MAGIC (open circles). The $Swift$-XRT light curve points are also plotted as grey points. The periodicity in the source is clearly evident in both TeV and X-ray energies. }
  \label{Fig3}
 \end{figure} \section{HESS J0632+057}
 
 	HESS J0632+057 was first detected as a TeV source serendipitously by the HESS collaboration in 2007 \cite{bib:HESSJ0632}. Its proximity to the Be star MWC 148 indicated that HESS J0632+057 could possibly be related to a binary system, similar to the Be star binary system associated with LS I +61$^{\circ}$ 303. When VERITAS observations of the source from 2006-2009 only resulted in upper limits (well below the HESS detected flux), it was clear that the source was not only variable, but most likely a new TeV binary source \cite{bib:GernotHESS, bib:Hinton2009}. $Swift$-XRT studies of HESS J0632+057 revealed a periodicity in the hard X-ray regime of 321$\pm$5 days \cite{bib:Bongiornio2011}, with optical spectroscopy using this periodicity to define the orbital eccentricity (e=0.83$\pm$0.08), periastron passage ($\phi$=0.97), and apastron passage ($\phi$=0.47) of the compact object (unknown nature) around MWC 148 \cite{bib:CasaresJ0632}. It should be noted that these derived orbital parameters have an uncertainty of 30-50$\%$. A zero transform discrete correlation function analysis of the \textit{Swift}-XRT dataset yields a modulation of 315$^{+6}_{-4}$ days, consistent with the results of \cite{bib:Bongiornio2011}.  This identification of HESS J0632+057 as a binary system marks the first time that such a system has been initially identified from TeV gamma-ray observations, a testament to the growing impact of the field of TeV astronomy. 
 
   \begin{figure}[t]
  \centering
  \includegraphics[width=0.5\textwidth, height=0.23\textheight]{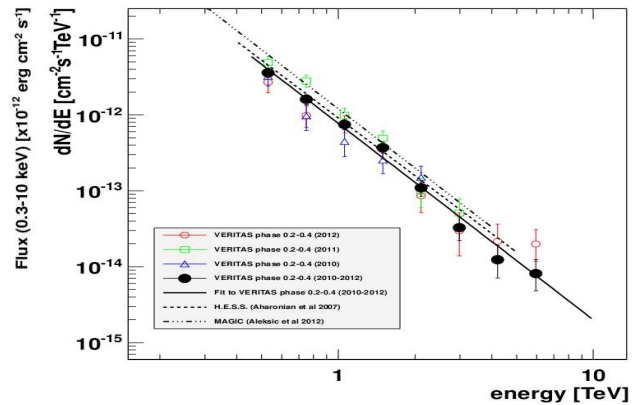}
  \caption{The differential energy spectrum from VERITAS, HESS, and MAGIC TeV observations taken from phases 0.2-0.4 only. The source is remarkably stable during its active phases, showing only variations in the flux normalization.}
  \label{Fig4}
 \end{figure}
 
 As of 2013, VERITAS has accrued over 160 hours on HESS J0632+057,  detecting the source near apastron (phases $\phi$=0.2-0.4) in 2010-2013 each year, yielding a total combined detection significance of 15.5$\sigma$ for emission $\>$1 TeV.  When active, the source demonstrates additional variability with flux $\>$1 TeV ranging from 2-5$\%$ of the Crab Nebula flux over the energy range. The source was particularly active in 2011 when TeV observations with VERITAS, HESS, and MAGIC were triggered by $Swift$-XRT observing a sharp increase in the 0.3-10 keV by a factor of $\sim$3 \cite{bib:ATELS1, bib:ATELS2, bib:MAGICJ0632, bib:VERITASandHESS}.  Combining the VERITAS, HESS, and MAGIC observations of HESS J0632+057 yields an 9 year long dataset on the source,  allowing for detailed examination of the TeV emission as a function of the 321 day period. When the complete TeV dataset is binned by orbital phase (Figure 4), the TeV emission appears to be consistent with the modulation seen in X-ray observations. The energy spectrum derived from the TeV detections (Figure 5) shows remarkable stability from orbit to orbit, with only the flux normalization changing over time and not the spectral index. 

 The similarities between the TeV emission behavior seen in HESS J0632+057 and LS I +61$^{\circ}$ 303 are certainly worth noting: they both only tend to emit TeV gamma-rays near apastron passage of their compact objects and the spectral index of the source appears to be fixed over many different orbits. However, HESS J0632+057 appears to be a much more stable TeV source with respect to its flux vs orbital phase than LS I +61$^{\circ}$ 303, which was not detected over many orbits for a time span of two years \cite{bib:USLSI3}. Additionally, HESS J0632+057 has \textit{not} been detected by $Fermi$-LAT in the GeV band despite dedicated searches \cite{bib:GeVJ0632}. This discrepancy could mean that the two sources are fundamentally different types of binary objects (i.e. micro quasars vs binary pulsar) or it could imply that the emission geometry is simply different between the two systems. Additional observations over many upcoming orbital cycles will help to disentangle the differences between the two sources and perhaps add key insights to what is powering both of the these objects.

\section{Conclusions}

The VERITAS TeV binary program has provided in-depth characterization of the TeV binaries LS I +61$^{\circ}$ 303 and HESS J0632+057. Despite extensive observations, it is still unclear as to what the fundamental properties that define active TeV binaries are (such as LS I +61$^{\circ}$ 303, HESS J0632+057, LS 5039, and PSR B1259-63), as opposed to other HMXBs that have not been detected at TeV energies such as Cygnus X-3 and GRS 1915+105. 

PSR B1259-63 is the only system in the class of TeV binaries with a known compact object (pulsar) and it is thought that the emission in PSR B1259-63 is the result of particles accelerated at the shock front between the pulsar and stellar wind of the Be star. HESS and $Fermi$-LAT observations of a flaring state in PSR B1259-63 \cite{bib:ISusch} show a similar gap in emission between the GeV and TeV energy ranges as the gap in LS I +61$^{\circ}$ 303 presented in this work. Given this, it might be reasonable to assume that similar emission mechanisms might be at work in both systems (i.e. pulsar wind powered). In Figure 6 are shown the relative distances of the orbits of the known TeV binary systems as well as the orbital phases during which they have been observed to be strong and weak sources. It is clear that even at periastron passage, PSR B1259-63 is still significantly further away from its companion star than LS I +61$^{\circ}$ is during its \textit{apastron} passage. If the same (pulsar driven) mechanisms are at work in LS I +61$^{\circ}$ 303 as in PSR B1259 -63, then why do the two systems tend to emit TeV radiation at completely different regions of their orbits? 

Additionally, in the wind-wind interaction model for emission, the density and velocity of the main sequence star wind plays a key role; why then, are the relative distances between the compact object and companion star so different during periods of activity from PSR B 1259-63 and LS I +61$^{\circ}$ 303? Taking into account the increased magnetic field at the shock during regions of high stellar wind density, it could be that the density of the stellar wind has to be \textit{less} than a critical amount in order for the accelerated particles at the shock to not suffer synchrotron losses that prevent them from being accelerated to TeV energies. This would explain the similarities between the emission properties of LS I +61$^{\circ}$ 303 and PSR B1259-63. However, this would also imply that LS I +61$^{\circ}$ 303 should be a constant TeV source away from periastron, instead of only select phases of its apastron passage.

From Figure 6, HESS J0632+057 and LS I +61$^{\circ}$ 303 seem to display strong emission near similar regions of their apastron passage, as well as weak emission again later in the orbit nearing periastron. This might imply that these two targets are similar in nature and in mechanisms of particle acceleration. However, HESS J0632+057 has still not been detected by $Fermi$-LAT, a key difference between the two sources. It could be that HESS J0632+057 is a weaker source and presents a GeV flux undetectable by $Fermi$-LAT, which seems to be supported by radio and X-ray observations \cite{bib:Skilton}. 

It is still unclear what the fundamental similarities are between the known TeV binary systems; further observations with VERITAS, HESS-II, MAGIC-II and the future CTA array will allow for more in depth investigation of this fascinating source class.

    \begin{figure}[t]
  \centering
  \includegraphics[width=0.52\textwidth]{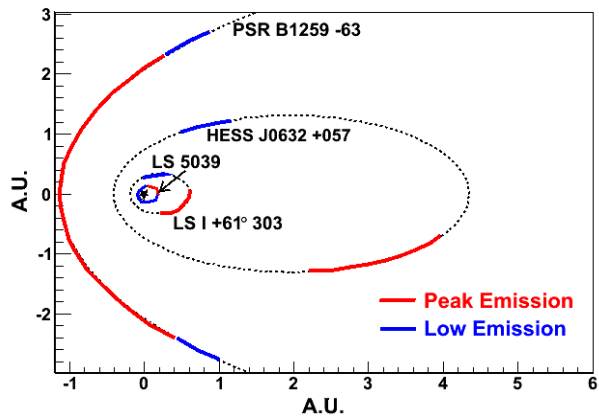}
  \caption{The relative orbital distances of the known TeV binaries in astronomical units. The orbits are shown without respect to inclination or viewing angle. Distances derived from \cite{bib:CasaresJ0632, bib:Aragona2009, bib:ISusch}. The colored line represent regions of higher and lower TeV activity from the systems.}
  \label{Fig3}
 \end{figure}

\vspace*{0.1 cm}
\footnotesize{{\bf Acknowledgment:}{This research is supported by grants from the U.S. Department of Energy Office of Science, the U.S. National Science Foundation and the Smithsonian Institution, by NSERC in Canada, by Science Foundation Ireland (SFI 10/RFP/AST2748) and by STFC in the U.K. We acknowledge the excellent work of the technical support staff at the Fred Lawrence Whipple Observatory and at the collaborating institutions in the construction and operation of the instrument.}}

\end{document}